\documentclass[aps,twocolumn,amsmath,amssymb,pra,showpacs,floatfix]{revtex4-1}
\usepackage{graphicx}
\usepackage{color}
\usepackage{bbold}
\usepackage{dcolumn}
\usepackage{bm}
\setcitestyle{numbers,square}

\begin{document}

\title{Bulk and edge-state arcs\\ in non-hermitian coupled-resonator arrays}

\author{Simon Malzard}
\affiliation{Department of Physics, Lancaster University, Lancaster, LA1 4YB, United Kingdom}

\author{Henning  Schomerus}
\affiliation{Department of Physics, Lancaster University, Lancaster, LA1 4YB, United Kingdom}

\begin{abstract}
We describe the formation of bulk and edge arcs in the  dispersion relation of two-dimensional coupled-resonator arrays that are topologically trivial in the hermitian limit. Each resonator provides two asymmetrically coupled internal modes, as realized in noncircular open geometries, which enables the system to exhibit non-hermitian physics. Neighboring resonators are coupled chirally to induce non-hermitian symmetries. The bulk dispersion displays Fermi arcs connecting spectral singularities known as exceptional points, and can be tuned to display purely real and imaginary branches.
At an interface between resonators of different shape,
one-dimensional edge states form that spectrally align along complex arcs connecting different parts of the bulk bands.
We also describe conditions under which the edge-state arcs are free standing.
These features can be controlled via anisotropy in the resonator couplings.
\end{abstract}
\maketitle

\section{Introduction}

Systems exhibiting non-trivial band topology attract intense attention
owing to phenomena such as chiral edge and surface states supported by boundaries between topologically distinct gapped phases \cite{RevModPhys.82.3045,Bansil2016}.
While the initial focus was on electronic or superconducting systems, the excitement quickly extended to other arenas of quantum and classical wave phenomena, which often display very different constraints than fermionic systems. In particular, the particle number is typically nonconserved  in classical \cite{Hafezi2011,Fang2012} and quantum  \cite{PhysRevX.6.041026} optical realizations, as well as for analogous emergent bosonic excitations such as polaritons \cite{PhysRevX.5.011034,PhysRevX.5.031001,St-Jean2017,PhysRevLett.120.097401}, phonons \cite{Suesstrunk47,PhysRevLett.114.114301,PhysRevX.5.031011}, and magnons \cite{PhysRevB.87.174427}.
This provides a bridge to non-hermitian systems that come along with their own complementary set of symmetries, as pioneered
with the realization that parity-time ($\mathcal{PT}$) symmetric non-hermitian Hamiltonians can display a real spectrum \cite{Bender1998}.
In this and other non-hermitian settings, the band-degeneracy points that play a crucial role at separating topologically distinct gapped phases are replaced by exceptional points (EPs) \cite{Berry2004,Heiss2012,Rotter2009}, where
eigenvalues depart from spectral symmetry lines \cite{Bender2007}. Just as their hermitian counterpart  (known in general as diabolic points, DPs), EPs carry a topological charge derived from a geometric phase,
which prevents their spontaneous isolated creation or annihilation unless two points of opposite charge collide \cite{Berry2004}.

In recent years, several mechanisms have been reported to induce topological phenomena into non-hermitian systems, be it based on non-hermitian time-reversal symmetries (as for $\mathcal{PT}$) or analogous combinations with chiral  symmetry ($\mathcal{X}$) or charge-conjugation symmetry ($\mathcal{C}=\mathcal{XT}$). A main motivation is to obtain topological states with distinct life times, as encoded in the imaginary part of the energy spectrum.
In many cases, the models are based on topological hermitian counterparts, such as the well studied Su-Schrieffer-Heeger (SSH) model \cite{SSH1979}, where dissipation can yield quantized displacements of decay and survival processes \cite{Rudner2009,Rudner2016}, while the
introduction of gain and loss yields a topological mechanism of zero-mode selection based on a nonhermitian $\mathcal{C}$  \cite{Schomerus2013,Poli2015,Ge2017,Zhao2018,Parto2018} or  $\mathcal{PT}$ \cite{Weimann2017} symmetry. In these and other examples \cite{Mahito2011,Zhao2015} derived from topological hermitian systems, the topologically protected states still obey a bulk-boundary correspondence \cite{Leykam2017} and their robust properties are directly inherited from the hermitian limit.

However, this is not always the case. Protected edge and interface states can also arise via EPs, even when the hermitian limit is topologically trivial \cite{Malzard2015,Lee2016,Yao2018,Pan2018}. This mechanism equips a system with robust spatially localized states that display distinct life times.
Furthermore, non-hermitian effects can fundamentally change the properties of edge states of a hermitian origin, which for instance can bifurcate at EPs to display additional branches with $\mathcal{PT}$-symmetry
\cite{Yidong2018}.
These observations highlight the role that EPs and their topological charges play in
distinguishing conventional topological states with adopted non-hermitian properties from genuinely non-hermitian symmetry-protected states that do not have a hermitian counterpart  \cite{Leykam2017,Shen2017}, and for classifying nonhermitian topological systems in general \cite{Gong2018}.

\begin{figure*}[t]
\includegraphics[width=.95\textwidth]{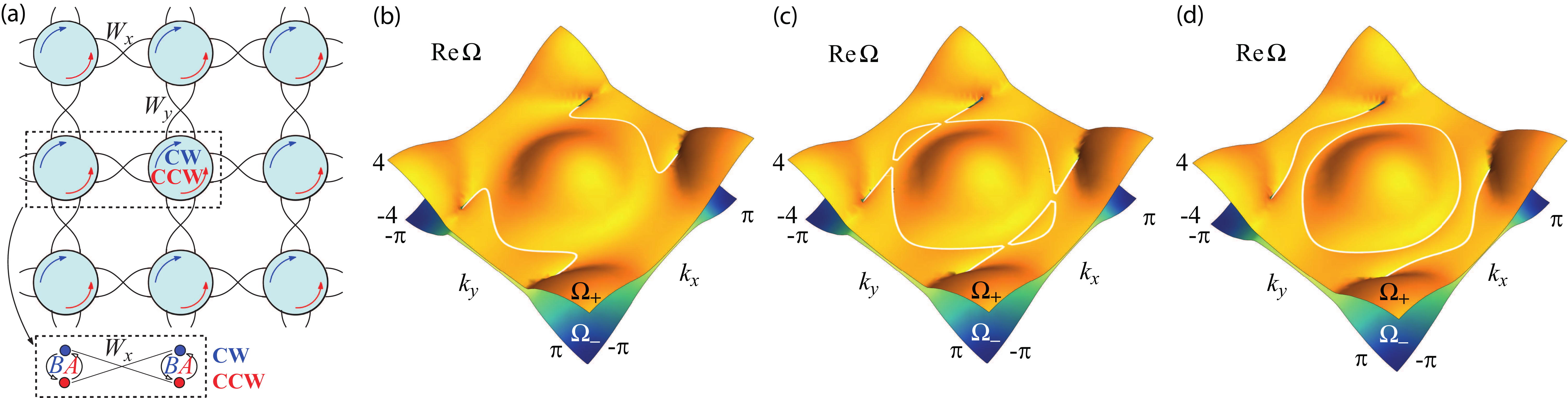}
\caption{Bulk Fermi arcs in a two dimensional array of evanescently coupled nonhermitian resonators. (a) Each resonator supports a clockwise (CW) and a counter-clockwise (CCW) internal mode that are coupled by asymmetric backscattering amplitudes $A$ and $B$, as obtained, e.g., from a small nonspherical deformation of open dielectric resonators. The resonators are placed on a square lattice and are coupled evanescently with coupling coefficients $W_x$ and $W_y$ that convert CW waves into CCW waves. This coupling configuration introduces a chiral symmetry into this nonhermitian system.
(b-d) Real part $\mathrm{Re}\,\Omega$ of the bulk dispersion for $B=-2.5+0.2i$, $W_{x}=1.0+0.1i$, $W_{y}=1.0+0.5i$, and the three values $A=1.5+0.1i$ (b), $A=1.5+0.2i$ (c), and $A=1.5+0.3i$ (d). In each case, the dispersion consists of two sheets $\Omega_+$ (yellow surface) and $\Omega_-=-\Omega_+$ (blue surface) that are related by the chiral symmetry. The white lines indicate Fermi arcs and lines with $\mathrm{Re}\,\Omega=0$, corresponding to intersections of the two sheets. The arcs terminate at exceptional points (EPs), which are the nonhermitian counterparts of Weyl points in topological insulators. In (b), four EPs are connected by two arcs. In (d), the EPs are reconnected by two arcs with a different topology, while a closed Fermi line is also present. Panel (c) shows the reconnection point between these two scenarios, which is mediated by two smaller closed Fermi lines.}
\label{fig1}
\end{figure*}

Spectral singularities also play a role in non-gapped topological systems, such as three-dimensional Weyl and Dirac semimetals where such singularities appear at generic DPs in momentum space \cite{NIELSEN1983389,PhysRevLett.107.127205}.
A particularly intriguing
phenomenon observed in such systems is the formation of surface-state Fermi arcs,
corresponding to dispersive branches in the band structure that connect topologically distinct parts of the bulk bands and are hinged to these spectral singularities \cite{PhysRevB.83.205101,Xu613}.
In  optical systems, hermitian Fermi arcs have been
realized in metamaterials \cite{Yang2017} and laser-written waveguides \cite{Yidong2017}. This provides again a springboard to explore nonhermitian counterparts.
EPs occur generically when only two parameters are varied in a non-hermitian system \cite{Berry2004}, while a non-hermitian perturbation of a DP in three dimensions results in a one-dimensional exceptional curve \cite{PhysRevB.97.075128}.
Incorporating loss into a two-dimensional topological system with a DP therefore results in the formation of two EPs  connected by a bulk Fermi arc, a scenario that has been realized in a periodic photonic crystal \cite{Zhou2018}.
Nonhermiticity can also close the band gap in the bulk of a system, resulting in the formation of exceptional points and curves in momentum space that spawn complex branches in the bulk dispersion \cite{PhysRevA.84.021806,PhysRevA.85.013818}.
Furthermore, the $\mathcal{PT}$-symmetric edge states branching off conventional edge states \cite{Yidong2018} have been found to survive when a band gap has been closed in such a way.

These complex spectral bulk and edge effects in gapless topological systems provide the backdrop for the present paper.
The key question which we pursue in this work is whether analogous effects  can also be achieved in systems with a topologically trivial hermitian limit. To this end, we show that one can indeed realize bulk Fermi arcs and complex edge-state arcs in a highly accessible setting, given by a two-dimensional array of microresonators.
Each individual resonator is taken to be open and deformed into a
slightly non-circular shape, a combination of features that
has been exploited widely both in theory and experiment to induce nonhermitian effects akin to $\mathcal{PT}$ symmetry \cite{RevModPhys.87.61}.
The array places these resonators onto a two-dimensional square lattice where neighboring resonators are coupled evanescently. This gives rise to a
chiral symmetry, as previously employed in photonic analogues of topological insulators \cite{Hafezi2011,PhysRevLett.110.203904}. In our setting, however, the system belongs to
a universality class (the chiral universality class AIII) which for hermitian two-dimensional systems is topologically trivial \cite{1367-2630-12-6-065010,RevModPhys.82.3045}.
By adjusting a single parameter, the array can  furthermore be tuned to jointly display $\mathcal{PT}$ and $\mathcal{C}$ symmetry. In the general classification of topological systems,
this modification places the system into a different symmetry class (the chiral class with conventional time-reversal symmetry, BDI),
which however is again topologically trivial in the hermitian limit.

In the symmetry class AIII, we demonstrate the existence of bulk Fermi arcs that emanate from EPs
in the Brillouin zone.
In the symmetry class BDI, the symmetries combine to stabilize purely real and imaginary branches
of the bulk dispersion.
These bulk effects are described along with the details of the model in Section \ref{sec:bulk}.

To induce edge effects into the system we join two arrays with opposite resonator deformations along an interface.
In one-dimensional coupled-resonator optical waveguides, such an interface can give rise
to defect states, which emerge at EPs that are passed when the system is sufficiently non-hermitian \cite{Malzard2015}.
In the two-dimensional array, this effect takes place in momentum space, resulting in complex edge-state arcs. The arcs depart from the bulk bands, and provide states with distinct mode profiles, frequencies and life times. These edge effects are described in Section \ref{sec:edge}.

The spectral phenomena described in this work are highly adaptable via the choice of the resonator geometry and the resonator placement. In particular, the evanescent coupling strengths between the resonators can be used to reposition the exceptional points, and also
control which parts of the edge-state arcs are physically accessible in the Brillouin zone. These couplings therefore
provide a simple mechanism to tailor and accentuate the predicted spectral features of the array.

\section{Model and bulk band structure}
\label{sec:bulk}

The two-dimensional resonator array is sketched in Fig.~\ref{fig1}(a). It consists of dielectric resonators that
confine light by internal reflection, resulting in resonance states of finite life captured by complex resonance frequencies $\Omega$. We consider the generic effects of openness and noncircular deformation of these resonators, in a regime where optical reciprocity is maintained and the coupling of different resonators is predominantly evanescent.

In circular resonators, one encounters spectrally degenerate but
mutually orthogonal pairs of clockwise (CW) and counter-clockwise (CCW) whispering-gallery modes.
Even when the life times of these modes are large,
strong non-hermitian effects ensue when such resonators
are slightly but asymmetrically deformed, be it smoothly or via notches or attachment of nanoparticles \cite{PhysRevA.84.023845,RevModPhys.87.61}. We consider a spectral range in which only one of these pairs  in each resonator is relevant, and denote the amplitude of the CW mode  $a_{n,m}$ and the amplitude of the CCW mode as $b_{n,m}$, where the resonator location is identified by the indices $n,m$ pertaining to the $x,y$ directions. The amplitudes in the resonator are placed together into a two-component vector $\boldsymbol{\psi}_{n,m}={a_{n,m}\choose b_{n,m}}$.
The two-mode Hamiltonian describing an individual isolated resonator then has the form
\begin{equation}
h =
\begin{pmatrix}
\Omega_{0} & A\\
B & \Omega_{0}
\end{pmatrix}.
\label{eq:h}
\end{equation}
As detailed in  Appendix \ref{appA}, reciprocity dictates that the whispering-gallery modes retain an identical bare resonance frequency $\Omega_{0}$. The main effect of the deformation is a finite but asymmetric (directed) backscattering between both modes, with an amplitude $A$
for CCW to CW conversion that can substantially differ in magnitude and phase from the amplitude $B$ for CW to CCW conversion \cite{PhysRevA.84.023845,RevModPhys.87.61}. This asymmetry is intimately related to the openness of the system, and results in the splitting of CW/CCW hybridized resonance states, which have complex frequency $\Omega_0\pm\sqrt{AB}$. In the following, we refer all frequencies to the references frequency $\Omega_0$, thereby effectively setting $\Omega_0=0$.

In the array, adjacent resonators are coupled together evanescently, such that CW modes from one resonator couple to the CCW modes of an adjacent resonator. We can safely assume that inter-resonator backscattering (CW to CW and CCW to CCW) is suppressed, which is well fulfilled if the  approaching resonator boundaries are smooth on the scale of the wavelength  \cite{Hafezi2011,PhysRevLett.110.203904,PhysRevA.90.053819}.
However, we allow that the coupling strengths for resonators adjacent along the $x$ direction is different from the coupling strengths for resonators adjacent along the $y$ direction.
This corresponds to
coupling matrices (for a detailed justification and generalizations see Appendix \ref{appA})
\begin{equation}
t_x =
\begin{pmatrix}
0 & W_x\\
W_x & 0
\end{pmatrix},\quad
t_y =
\begin{pmatrix}
0 & W_y\\
W_y & 0
\end{pmatrix}.
\label{eq:w}
\end{equation}
With these definitions, the coupled-mode equations for a uniform two-dimensional array are given by
\begin{eqnarray}
\Omega\boldsymbol{\psi}_{n,m}=h\boldsymbol{\psi}_{n,m} &&+ t_{x}(\boldsymbol{\psi}_{n+1,m} +  \boldsymbol{\psi}_{n-1,m}) \nonumber \\ &&+ t_{y}(\boldsymbol{\psi}_{n,m+1} +  \boldsymbol{\psi}_{n,m-1}).
\label{eq:cmeqs}
\end{eqnarray}

Stipulating that the solutions are of the form $\boldsymbol{\psi}_{n,m}=e^{ik_{x}n + ik_{y}m}{a\choose b}$, we obtain the corresponding Bloch Hamiltonian
\begin{equation}
H(\mathbf{k})=
\begin{pmatrix}
0 & A+2W_{x}C_{x}+2W_{y}C_{y} \\
B+2W_{x}C_{x}+2W_{y}C_{y} & 0
\end{pmatrix},
\label{BlochH}
\end{equation}
where $\mathbf{k}={k_x\choose k_y}$, $C_{x}=\cos{k_{x}}$ and $C_{y}=\cos{k_{y}}$.
 The eigenvalues of $H(k)$ provide the dispersion relation
\begin{align}
&\Omega_{\pm}(\mathbf{k}) =
\pm \sqrt{(A+2W_{x}C_{x}+2W_{y}C_{y})(B+2W_{x}C_{x}+2W_{y}C_{y})}.
\label{bloch}
\end{align}

In general, the parameters $A$, $B$, $W_x$, and $W_y$ are complex, so that time-reversal symmetry is broken. In this case, the Bloch Hamiltonian still displays a chiral symmetry $\sigma_z H(\mathbf{k})\sigma_z =-H(\mathbf{k})$ with Pauli matrix $\sigma_z$, according to which the two eigenvalues are constrained to $\Omega_-=-\Omega_+$ as indeed observed above.
In the general classification of hermitian topological systems, this places the systems into symmetry class AIII, which is topologically trivial in two dimensions.

As shown in the examples of Fig.~\ref{fig1}(b-d), representative band structures in this class AIII combine regions with predominantly real and predominantly imaginary resonance frequencies. In all cases, one can clearly make out bulk Fermi arcs with $\mathrm{Re}\,\Omega_{\pm}=0$. These arcs emanate from EPs, which arise when
\begin{subequations}\label{eq:bulkeps}
\begin{eqnarray}
&& A+2W_{x}C_{x}+2W_{y}C_{y}=0\quad\mbox{or}
\\
&&B+2W_{x}C_{x}+2W_{y}C_{y}=0.
\end{eqnarray}%
\end{subequations}
Each of these conditions can be met by varying two real parameters such as $k_x$ and $k_y$, so that the EPs appear generically at isolated positions in the two-dimensional Brillouin zone. The arcs can occur on their own [see panel (b)]
or be complemented by closed Fermi lines (d).
The topology of these lines and arcs can change at parameters for which they intersect (c), which
occurs when an arc crosses a stationary point $\partial_{k_x+ik_y}\Omega_\pm|_{k_x-ik_y}=0$ or $\partial_{k_x-ik_y}\Omega_\pm|_{k_x+ik_y}=0$.

In order to better understand these features, we tune the array further to display a non-hermitian $\mathcal{PT}$ symmetry. This is obtained when in a suitable basis all parameters $A$, $B$, $W_x$ and $W_y$ are real. Real values of $W_x$ and $W_y$ are realized when the evanescent coupling is lossless \cite{PhysRevLett.113.123903}. To obtain real but distinct values of $A$ and $B$, we need to keep the resonators open but only need to tune a single parameter \cite{PhysRevA.90.053819}, as the relative phase of these amplitudes can be adjusted by choice of the whispering-gallery mode basis (see Appendix \ref{appA}).
The hybridized resonance modes in each isolated resonator then have complex frequencies $\Omega_0\pm\sqrt{AB}$, so that the resonance splitting is either purely real (if $AB>0$) or purely imaginary (if $AB<0$). In the general classification of hermitian topological systems, the case of real couplings represents the chiral class BDI with a conventional time-reversal symmetry (here obtained from $(\mathcal{PT})^2=1$), which for two-dimensional  systems is again topologically trivial.

Under these conditions,
the band structure \eqref{bloch}
of the resonator array is real and gapped if $|A|,|B|>2(W_{x}+W_{y})$ and $AB>0$, or imaginary and gapped if for the same conditions $AB<0$. In all other cases, the dispersion contains purely real and purely imaginary branches in ranges of $k_{x}$  and $k_{y}$. These branches are joined at lines of EPs with $\Omega_{\pm}=0$, which are again determined by the condition \eqref{eq:bulkeps}.

These additional spectral constraints are consequences of the manifest symmetries of the system in the CW/CCW basis. The nonhermitian $\mathcal{PT}$ symmetry makes the Bloch Hamiltonian \eqref{BlochH}
real, $H(\mathbf{k})=H^*(\mathbf{k})$, so that its eigenvalues are either real or form a complex-conjugated pair. The combination with the chiral symmetry $\sigma_z H(\mathbf{k})\sigma_z =-H(\mathbf{k})$ yields the $\mathcal{C}$ symmetry $\sigma_z H(\mathbf{k})\sigma_z =-H^*(\mathbf{k})$.
From this we find that the complex eigenvalues in a conjugate pair must both obey $\Omega_\pm=-\Omega_\pm^*$, hence must both be purely imaginary.

\begin{figure}
\includegraphics[width=\columnwidth]{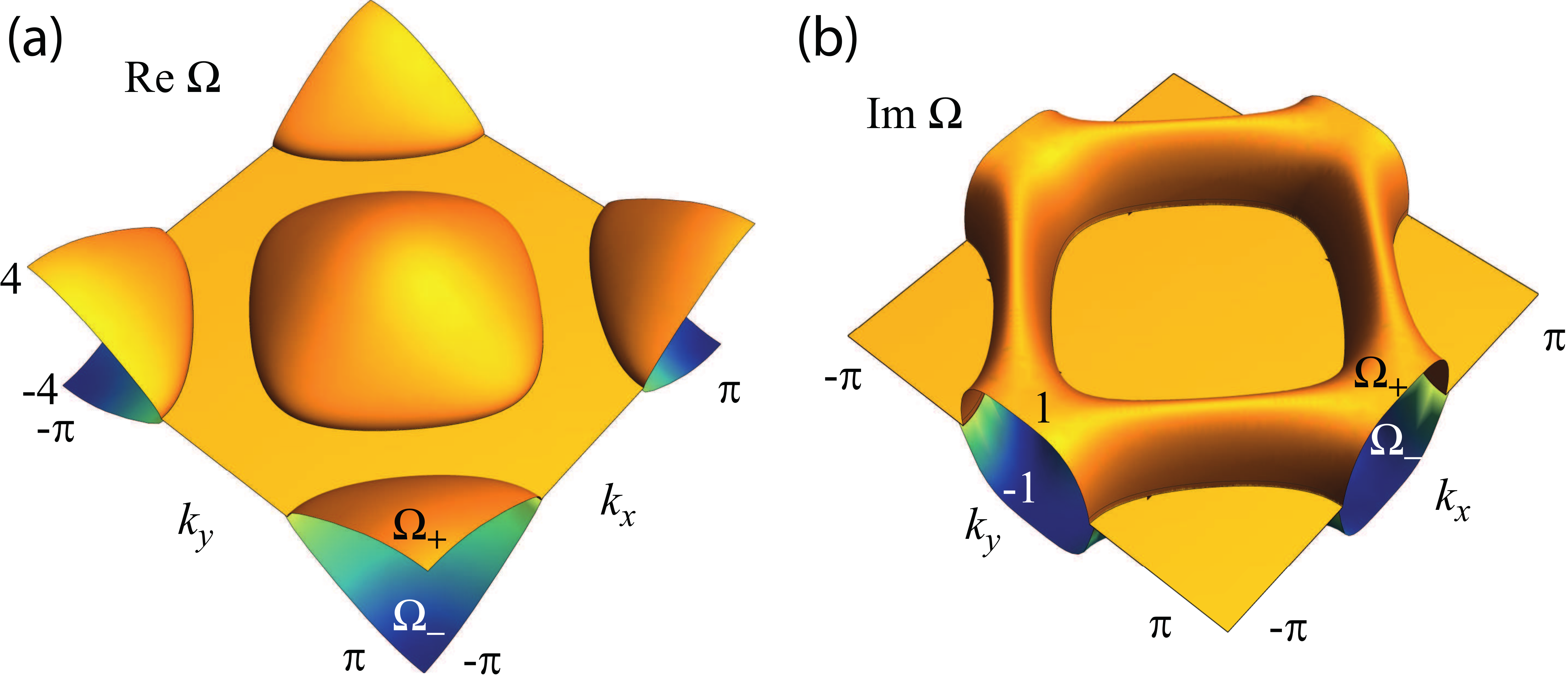}
\caption{(a) Real and (b) imaginary parts of the bulk dispersion for $A=1.0$, $B=-1.0$ and $W_{x}=W_{y}=1$, representing the $\mathcal{PT}$-symmetric case (symmetry class BDI) where the band structure displays purely real and imaginary branches, and the exceptional points degenerate into lines.}
\label{fig2}
\end{figure}

A representative example of a band structures in symmetry class BDI is shown in Fig.~\ref{fig2}.
The purely imaginary branches define flat patches with $\mathrm{Re}\,\Omega_\pm=0$, which are bounded by  lines of EPs. The Fermi arcs in symmetry class AIII can be interpreted as remnants of these regions when the $\mathcal{PT}$-symmetry is explicitly broken by complex coupling values.

While this exhausts all possibilities for the two-component Bloch Hamiltonian, the spectral constraints from $\mathcal{PT}$ symmetry can also, e.g., be fulfilled by quadruplets of eigenvalues $\Omega_1=-\Omega_2=\Omega_3^*=-\Omega_4^*$, which are arranged in a four-fold symmetric pattern with respect to the real and imaginary axis.
As we will see in the next section, this much richer picture unfolds in systems with suitable interfaces, where it provides the avenue to the formation of complex edge-state arcs.

\begin{figure}[t]
\includegraphics[width=\columnwidth]{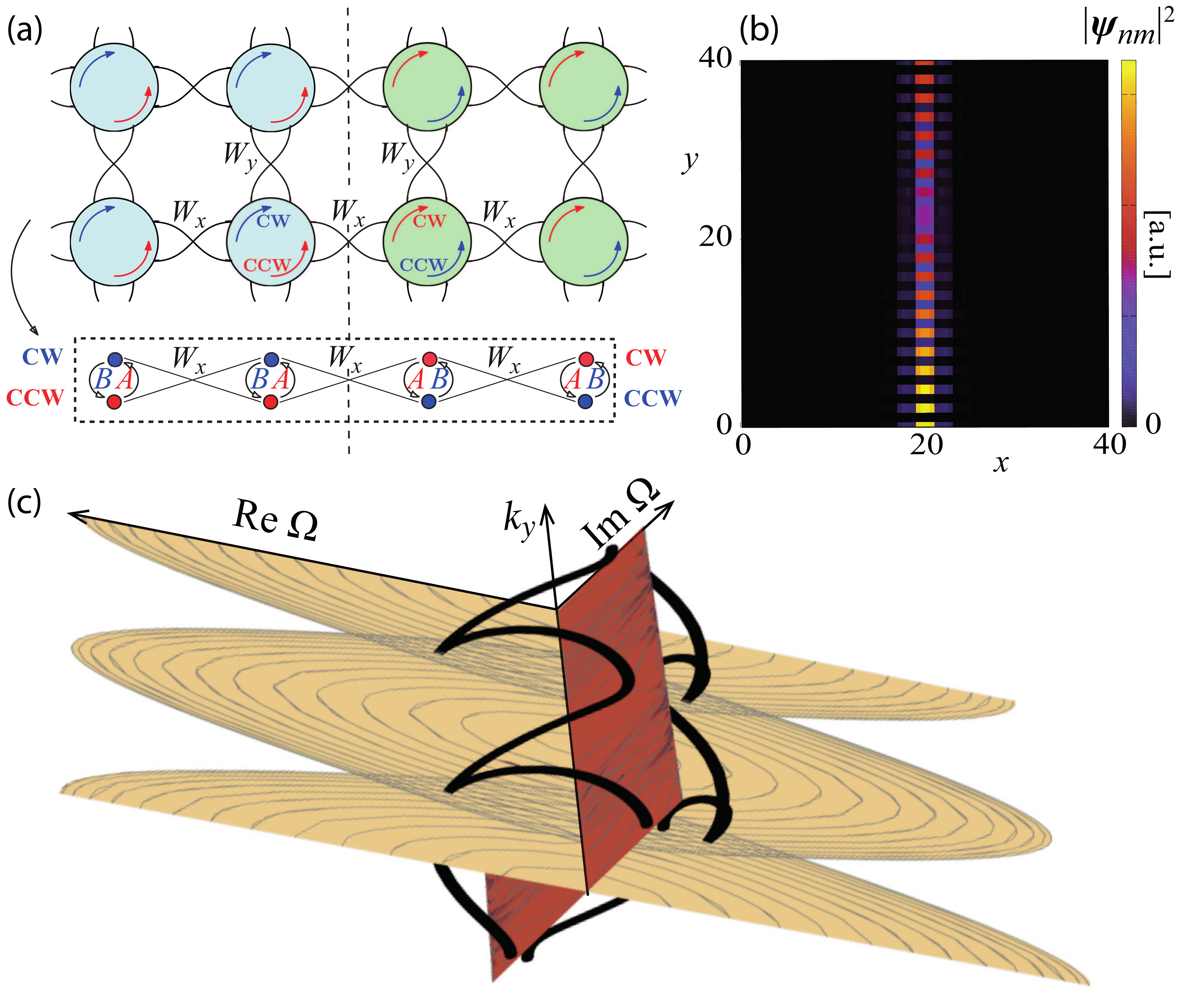}
\caption{Edge-state arcs in an array with  an interface joining resonator arrays with opposite backscattering.
(a) Horizontal slice through the array, where the dotted line indicates the interface between resonators with backscattering amplitudes $A$ and $B$ as in Fig.~\ref{fig1} (blue resonators to the left),
and resonators where the values of these backscattering amplitudes are interchanged (green resonators to the right). (b) Density plot of the intensity of a representative edge state in a finite square array of $40 \times 40$ resonators, with $A=-B=W_{x}=W_{y}$. (c) Quasi-one dimensional band structure in the infinite version of this array, where $k_y$ is a good quantum number. In this representation,
the bulk bands form sheets, which here lie in the real and imaginary plane as all parameters are real ($\mathcal{PT}$-symmetric symmetry class BDI, see Fig.~\ref{fig2}). The black curves are the edge-state arcs, which connect the different sheets.
}
\label{fig3}
\end{figure}

\section{Edge-state Arcs}
\label{sec:edge}

To create edge states in the array we modify the bulk configuration and create an interface between two regions representing opposite deformations of the resonators [see Fig.~\ref{fig3}(a)]. The interface is placed along the $y$ axis and separates a region of resonators with internal coupling coefficients $A$ and $B$ ($n<0$, resonators depicted in blue) from a region of reflected resonators for which these two values are swapped around ($n\geq 0$, resonators depicted in green).
Figure \ref{fig3}(b) shows an example of an edge state in an array of $40\times 40$ resonators with parameters $A=-2$, $B=2$, $W_{x}=W_{y}=1$, demonstrating that such states can indeed be formed. The states are exponentially confined in the direction away from the interface, and display a standing-wave pattern along the interface. As we show in the following, the complex frequencies of these states form arcs that connect different points on the bulk dispersion relation. This is illustrated for the given set of parameters in Fig.~\ref{fig3}(c), where the complex frequencies are shown as a function of the conserved quantum number $k_y$, assuming an infinite size of the array. In this representation, the extended states form continuous two-dimensional sheets, which represent the projection of the bulk dispersion \eqref{bloch}. The edge states align along one-dimensional curves that terminate on the sheets and represent the arcs in question.
We now unfold the general picture behind these observations.

\begin{figure}[th]
\includegraphics[width=.65\columnwidth]{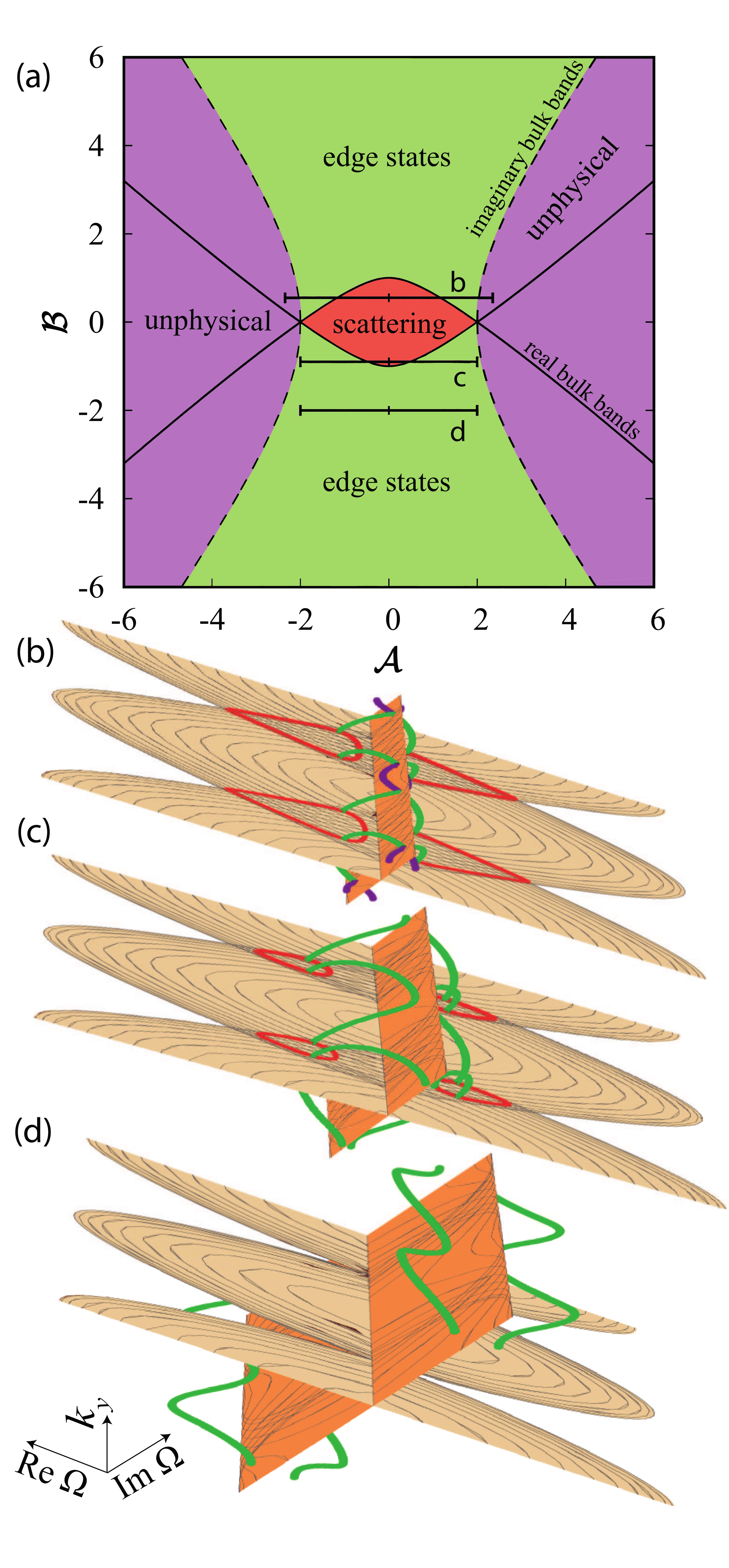}
\caption{(a) Traces of the edge-state arcs in the real section of their effective parameter space $(\mathcal{A},\mathcal{B})$ [defined in Eq.~\eqref{effective}]. The traces are horizontal lines of length $4W_y/W_x$, which are centred at $\mathcal{A}=(A+B)/2W_x$, $\mathcal{B}=(A-B)/2W_x$. The solid and dashed curves denote the termination conditions at the real and imaginary branches of the bulk bands, where the edge states (green region)  turn into extended scattering states (red) or into non-normalizable, unphysical states (blue).
The three representative traces correspond to the quasi-one-dimensional band structures shown in (b-d), where
the edge-state arcs are indicated in green, while their scattering predecessors are given in red and unphysical states in blue.
In (b), $A/W_x=0.55$, $B/W_x=-0.55$, $W_{y}/W_x=1.175$, for which the trace crosses both termination lines and the arcs connect the real and imaginary branches of the bulk bands.
In (c),  $A/W_x=-0.9$, $B/W_x=0.9$, $W_{y}/W_x=1$, for which the trace only reaches the real termination line so that the arcs loop back to the real branches. In (d), $A/W_x=-2.0$, $B/W_x=2.0$, $W_{y}/W_x=1$, for which the trace remains confined in the edge-state region so that the arc are free-standing.
}
\label{fig4}
\end{figure}

The translation symmetry along the $y$ axis allows us to separate variables according to
$\boldsymbol{\psi}_{n,m}=\boldsymbol{\varphi}_n e^{ik_{y}m}$, where the permitted values of $k_y$ are determined by the width of the array.
The wave equation \eqref{eq:cmeqs} then takes the form
\begin{equation}
\Omega\boldsymbol{\varphi}_{n}=h_{n}\boldsymbol{\varphi}_{n}+t_{x}(\boldsymbol{\varphi}_{n+1} + \boldsymbol{\varphi}_{n-1}),
\label{eq:cmeqs1d}
\end{equation}
where
\begin{equation}
h_n=\left\{
\begin{array}{ll}
        \begin{pmatrix}
0 & A^{\prime} \\
B^{\prime}  & 0
\end{pmatrix}
 & (n<0) \\[.3cm]
                \begin{pmatrix}
0 & B^{\prime} \\
A^{\prime}  & 0
\end{pmatrix} & (n\geq 0)
      \end{array}
    \right.
,
\label{eq:hn}
\end{equation}
with the effective coupling coefficients
\begin{subequations}\label{abprime}
\begin{eqnarray}
A^{\prime} = A+2W_{y}\cos{k_{y}}, \label{aprime}
 \\
B^{\prime} = B+2W_{y}\cos{k_{y}}.
\label{bprime}
\end{eqnarray}%
\end{subequations}

For fixed parameters $A',B'$, Eqs. \eqref{eq:cmeqs1d} and \eqref{eq:hn} define a quasi-one-dimensional set of coupled-mode equations,
which for $k_y=\pi/2$ recover the case of a one-dimensional array that admits complex defect states at sufficiently strong non-hermiticity \cite{Malzard2015}.
In this parameter space, the edge states can be obtained by wave matching, which is carried out in detail in Appendix \ref{appB}. Translated back into $k$ space, we obtain conditions determining the edge states which are conveniently expressed in terms of
the scaled frequency
\begin{equation}
\omega=\frac{\Omega}{W_x}
\end{equation}
and the effective parameters
\begin{subequations}\label{effective}%
\begin{align}
&\mathcal{A}=\frac{A+B+4W_y\cos k_y}{2W_x} \quad\quad\mbox{(arc parameter),}\\
&\mathcal{B}=\frac{A-B}{2W_x} \quad\quad\quad\mbox{(backscattering asymmetry)}.
\end{align}
\end{subequations}

Edge states with a symmetric mode profile about the interface then obey the equation
\begin{equation}
  -\omega\mathcal{A}^2+(2+\omega)\mathcal{B}^2+\omega(2-\omega)^{2}=0
,
\label{symmmain}
\end{equation}
while edge states with an antisymmetric mode profile obey the equation
\begin{equation}
\omega\mathcal{A}^2+(2-\omega)\mathcal{B}^2-\omega(2+\omega)^{2}
=0.
\label{antisymmmain}
\end{equation}
The symmetric and antisymmetric solutions are connected by chiral symmetry, thus remain paired as $\omega$ and $-\omega$ even when all parameters are complex (corresponding to symmetry class AIII).
In the $\mathcal{PT}$-symmetric case where the parameters $A$, $B$, $W_x$ and $W_y$ are real (symmetry class BDI), the solutions of Eqs.~\eqref{symmmain} and Eqs.~\eqref{antisymmmain} are further constrained to be real or to occur in complex-conjugated pairs, leading to quadruplets ($\omega,-\omega,\omega^*,-\omega^*$).

For a solution $\omega$ of Eq.~\eqref{symmmain} to define
a normalizable, physical edge state it furthermore has to fulfill the consistency condition
\begin{equation}
2-\lambda_{1}\lambda_{2}-\frac{1}{\lambda_{1}\lambda_{2}}=2\omega,
\label{eq:consistencymain}
\end{equation}
while the solutions of Eq.~\eqref{antisymmmain}
are subject to the consistency condition
\begin{equation}
2-\lambda_{1}\lambda_{2}-\frac{1}{\lambda_{1}\lambda_{2}}=-2\omega
\label{eq:consistency2main}.
\end{equation}
Here
\begin{subequations}\label{lambdamain}%
\begin{align}
&\lambda_{1}= C_{1}\pm \sqrt{C_1^2-1},\quad
C_1=
-\frac{\mathcal{A}}{2} + \frac{\sqrt{\mathcal{B}^2+\omega^{2}}}{2},\\
&\lambda_{2}= C_{2} \pm \sqrt{C_2^2-1},\quad
C_2=
-\frac{\mathcal{A}}{2} - \frac{\sqrt{\mathcal{B}^2+\omega^{2}}}{2},
\end{align}%
\end{subequations}
are propagation factors,
where the signs of the square roots have to be chosen such that $|\lambda_{1}|>1$ and $|\lambda_{2}|>1$.
If for a solution of Eqs.~\eqref{symmmain} or \eqref{antisymmmain}  a propagation factor is $|\lambda_{l}|=1$, we instead obtain a scattering state. If the consistency equations \eqref{eq:consistencymain} or \eqref{eq:consistency2main} can only be fulfilled by combining a decaying and an increasing propagation factor $\lambda_{l}$, the state cannot be normalized and is unphysical. Therefore, in both cases edge states change from physical to unphysical at points at which one of the propagation factors attains $|\lambda_{l}|=1$. This corresponds to propagating waves in the bulk system, and therefore occurs when an edge-state arc meets the bulk dispersion relation, where it then terminates. In the $\mathcal{PT}$-symmetric case (symmetry class BDI), the termination points coincide with degeneracies in edge-state quadruplets at real or imaginary frequencies, thus constituting EPs.

\begin{figure}[t]
\includegraphics[width=.5\textwidth]{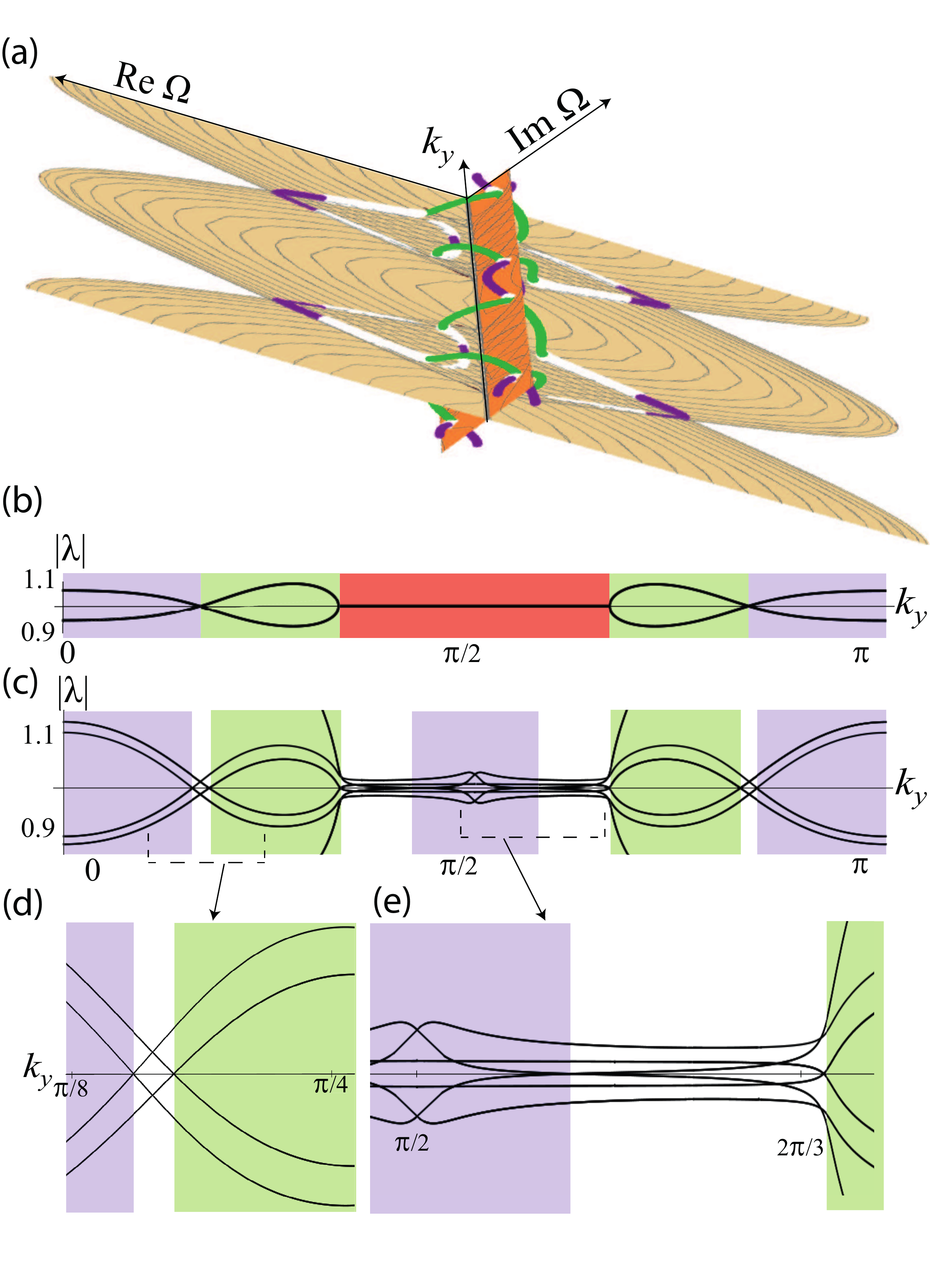}
\caption{(a)
Edge-state arcs for complex backscattering amplitudes $A/W_{x}=0.55+0.02i$, $B/W_{x}=-0.55+0.02i$ and $W_{y}=1.175$ [close to the real values in Fig.~\ref{fig4}(b)]. All arcs still terminate on the bulk bands, which now no longer are real or imaginary.
(b-d) Propagation factors $|\lambda_l|$, $1/|\lambda_l|$ of potential edge states as determined by Eq.~\eqref{lambdamain}. In (b),
 $A=-B=0.55/W_x$, $W_{y}/W_x=1.175$, corresponding to the real values of Fig.~\ref{fig4}(b).
In (c,d), the parameters take the complex values given above.
For complex parameters the region of scattering states is replaced by regions of physical and unphysical states. Furthermore, the termination points of different arcs now appear at separate values of $k_y$, as shown in detail in panels (d) and (e) which zoom into the termination region at the formerly purely imaginary and real branches of the bulk dispersion, respectively.}
\label{fig5}
\end{figure}

Figure \ref{fig4}(a) illustrates the traces of the edge-state arcs in  the real section of effective parameter space $(\mathcal{A},\mathcal{B})$, where $\mathcal{PT}$ symmetry holds.
The solid and dashed curves in the diagram denote the locations of the real and imaginary energy bands in this space (for analytical expressions see Appendix \ref{appB}),
and thereby confine regions in which the solutions of Eqs.~\eqref{symmmain} and  \eqref{antisymmmain}
represent genuine edge states (green region),
scattering states (red) or non-normalizable unphysical states (blue).
The edge states trace out a horizontal line $\mathcal{A}(k_y)$ in the arc parameter, which according to the definition \eqref{effective} has a length $4W_y/W_x$ determined by the coupling anisotropy, while its horizontal center is given by $\mathcal{A}(\pi/2)=(A+B)/2W_x$ and its vertical position given by the nonhermiticity parameter $\mathcal{B}=(A-B)/2W_x$.

The three labeled horizontal lines are the traces of edge states for representative systems with $A=-B$. In the complex energy dispersion these traces correspond to the arcs shown by the green curves in panels (b-d). These arcs can connect the real and imaginary branches of the bulk dispersion relation (b), can loop back to the real branch (c), or can be disconnected from the bulk bands (d).
The blue arcs in panel (b)  looping back to the imaginary branches represent unphysical states, arising in panel (a) from segments of the trace across the dashed line. The red arcs in panels (b,c) represent scattering states within the real branch of the bulk dispersion, which occur in panel (a) when the traces cross the corresponding solid line. The disconnected arcs as shown in panel (d) occur for traces that are confined to the interior of the edge-state region, so that they do not cross the phase boundaries defined by the bulk bands.

As shown in Fig.~\ref{fig5}, the edge-state arcs persist for complex coupling parameters, for which the $\mathcal{PT}$ symmetry is broken so that different arcs are only related by the chiral symmetry (corresponding to symmetry class AIII). In this figure, the chosen parameters are almost real, which allows us to identify the most important qualitative effects.
As seen in panel (a), the arcs still emerge from the bulk dispersion, but the degeneracy of the termination points is lifted, which is particularly visible on the formerly real sheets. Panels (b) and (c) compare the propagation factors $\lambda_l$ and $1/\lambda_1$ along the arc for real and complex couplings. This comparison reveals two distinct effects. The propagation factors of the former scattering states
acquire moduli $|\lambda_1|\neq 1$  and hence turn into weakly confined edge states or
non-normalizable, unphysical states. Furthermore, the formerly degenerate transitions at which edge states from peviously symmetry-paired arcs become unphysical occur at independent values of $k_y$,
as shown in panel (d) close to the formerly imaginary sheet and in panel (e) close to the formerly real sheet.
All these features remain dictated by the general quantization conditions
~\eqref{symmmain} and \eqref{antisymmmain} subject to the consistency equations \eqref{eq:consistencymain} and \eqref{eq:consistency2main}, which hold for general complex values of all coupling parameters.

\section{Conclusions}

In summary, two-dimensional resonator arrays with non-hermitian internal  backscattering and chiral coupling are capable of displaying a broad range of complex spectral phenomena in the bulk and at interfaces. In particular, the bulk dispersion relation of these systems can feature complex bulk Fermi arcs and flat patches (see Section \ref{sec:bulk} and Figs.~\ref{fig1}, \ref{fig2}), while the dispersion along an interface can exhibit complex edge-state arcs (see Section \ref{sec:edge} and Figs.~\ref{fig3}-\ref{fig5}).
The described effects replicate the behaviour previously predicted for systems whose hermitian limit is topologically nontrivial. However, the arrays studied here belong to universality classes that in the hermitian limit are topologically trivial.
The described set-up is highly flexible, so that the spectral effects are widely adaptable, where we here focussed on the main phenomenology.

The key feature from non-hermitian physics that enables these effects are exceptional points, which replace the spectral singularities in hermitian topological systems. These exceptional points can appear generically even in universality classes with a topologically trivial hermitian limit, as utilized here for two-dimensional reciprocal systems from class AIII (chiral symmetry) and BDI (chiral symmetry with a conventional time-reversal symmetry).
These properties further expand the scope of nontrivial physics that arises from nonhermiticity, as previously seen in the context of isolated topological defect states.

From a practical perspective, these findings imply that nontrivial dispersion effects can be achieved without needing to resort to carefully engineered systems that replicate the intricate symmetries required for hermitian topological physics. For instance, we assumed throughout that reciprocity is preserved. Therefore, the described effects can occur in conventional optical settings. The main requirement are chiral symmetry, as previously employed in photonic analogues of topological insulators \cite{Hafezi2011,PhysRevLett.110.203904}, and
asymmetric backscattering, which can be realized in a wide range of resonator geometries \cite{PhysRevA.84.023845,RevModPhys.87.61}.

In such optical settings, bulk Fermi arcs are directly observable  in  momentum space \cite{Zhou2018}. Focussing on signatures in the local and total density of states, they should also be observable, e.g., in microwave setups.
The edge-state arcs provide spatially localized states that do not exist in their hermitian counterpart.
Besides this characteristic spatial confinement, a key feature that distinguishes these states are their distinct life times. Therefore, an attractive approach to probe a system for edge-state arcs would be to excite a state locally near the interface, and observe the dynamical evolution of this state along the arc towards long life times. In this way, these resonator arrays provide promising mechanisms for state engineering that deserve further investigation.

\appendix

\section{Symmetry constraints}
\label{appA}
Time-reversal symmetry, reciprocity and chiral symmetries take different forms in different bases. Here we elucidate the resulting constraints in the standing-wave basis of appropriately normalized sine and  cosine waves $|s\rangle$, $|c\rangle$ around the perimeter of the resonators, and the whispering-gallery basis of clock-wise and counter-clockwise circulating modes
\begin{equation}
|\mathrm{CW}\rangle= \sqrt{\frac{1}{2}}(|c\rangle-i|s\rangle)e^{i\chi},\quad
|\mathrm{CCW}\rangle= \sqrt{\frac{1}{2}}(|c\rangle+i|s\rangle)e^{-i\chi}.
\label{eq:cwccw}
\end{equation}
The angle $\chi$ originates from the freedom to choose the nominal origin around the perimeter of the resonators. We initially set $\chi=0$ and discuss this freedom at the end of this Appendix.

Optical reciprocity requires that the non-hermitian effective  Hamiltonian is symmetric in the standing-wave basis. For an isolated resonator, this implies
\begin{align}
\hat h  =  &\alpha|c\rangle\langle c|+\beta(|c\rangle\langle s|+|s\rangle\langle c|)
+\gamma|s\rangle\langle s|
\\
 =&
\frac{\alpha+\gamma}{2}(|\mathrm{CW}\rangle\langle\mathrm{CW}|+|\mathrm{CCW}\rangle\langle\mathrm{CCW}|)
\nonumber \\ &
+(\frac{\alpha-\gamma}{2}+i\beta)|\mathrm{CW}\rangle\langle\mathrm{CCW}|
\nonumber \\ &+(\frac{\alpha-\gamma}{2}-i\beta)|\mathrm{CCW}\rangle\langle\mathrm{CW}|,
\end{align}
which justifies the form Eq.~\eqref{eq:h} of the resonator Hamiltonian $h$ with $\Omega_0=(\alpha+\gamma)/2$, $A=i\beta+(\alpha-\gamma)/2$ and $B=-i\beta+(\alpha-\gamma)/2$.

Let us now consider the coupling of two resonators, with the states in the second resonator denoted by a dash. The coupling operator then takes the form
\begin{align}
\hat t   =& W_{cc}(|c\rangle\langle c'|+|c'\rangle\langle c|)
+ W_{ss}(|s\rangle\langle s'|+|s'\rangle\langle s|)
\nonumber\\
+&W_{cs}(|c\rangle\langle s'|+|s'\rangle\langle c|)
+ W_{sc}(|s\rangle\langle c'|+|c'\rangle\langle s|)
\\
=&
W_+
(|\mathrm{CW}\rangle\langle\mathrm{CW}'|+|\mathrm{CCW}'\rangle\langle\mathrm{CCW}|)
\nonumber\\
+&W_-
(|\mathrm{CCW}\rangle\langle\mathrm{CCW}'|+|\mathrm{CW}'\rangle\langle\mathrm{CW}|)
\nonumber\\
+&W_+'
(|\mathrm{CW}\rangle\langle\mathrm{CCW}'|+|\mathrm{CW}'\rangle\langle\mathrm{CCW}|)
\nonumber\\
+&W_-'
(|\mathrm{CCW}\rangle\langle\mathrm{CW}'|+|\mathrm{CCW}'\rangle\langle\mathrm{CW}|),
\end{align}
with
\begin{align}
W_{\pm}=&\frac{W_{cc}+ W_{ss}\pm i (W_{sc}-W_{cs})}{2},\nonumber\\
W'_{\pm}=&\frac{W_{cc}- W_{ss}\pm i(W_{sc}+W_{cs})}{2}.
\end{align}

In the regime of evanescent coupling, backscattering is suppressed, which corresponds to
$W_{cc}+ W_{ss}\approx 0$, $W_{cs}-W_{sc}\approx 0$.
This then leads to a coupling matrix
\begin{equation}
t =
\begin{pmatrix}
0 & W'_+\\
W'_- & 0
\end{pmatrix}
\end{equation}
with $W'_+=W_{cc}+iW_{cs}$, $W'_-=W_{cc}-iW_{cs}$. Furthermore, evanescent coupling usually involves smooth, locally symmetric  geometries where also $W_{cs},W_{sc}\approx 0$, so that we set $W'_+= W'_-\equiv W$.
In the expressions
\eqref{eq:w} for $t_x$ and $t_y$, these coefficients are evaluated at the appropriate positions of adjacent resonators in the horizontal or vertical direction.

These conditions deliver the general model studied in our work, which always displays a chiral symmetry.
The additional $\mathcal{PT}$-symmetry emerges when we can find a basis in which all coupling coefficients are real. For the coefficients $W_x$ and $W_y$, this is fulfilled when the evanescent coupling does not induce any losses \cite{PhysRevLett.113.123903}.
For the internal couplings $A$, $B$, we argue that they can be made real by tuning a single generic parameter. To realize this, we exploit as mentioned above that in all these considerations we have a freedom to  choose the nominal origin of the standing or propagating waves along the perimeter of each resonator. For instance, rotating the origin in each resonator by an angle $\phi_0$, we can consider CW waves $\sim \exp(-im(\phi-\phi_0))$ and CCW waves $\sim \exp(im(\phi-\phi_0))$ with modes index $m$. In Eq.
\eqref{eq:cwccw}, this corresponds to an additional phase $\chi=m\phi_0$. The coupling amplitudes $A$ and $B$ then pick up opposite phase factors $A\to A \exp(-2i\chi)$, $B\to B \exp(2i\chi)$. Thus, to achieve $\mathcal{PT}$ symmetry  we only require to tune a single parameter to achieve that $A$ and $B$ have an opposite phase.
This phase can then be transformed to 0 by an appropriate choice of the propagating-wave basis. As a key indicator, one can consider the resonances $\Omega_0\pm\sqrt{AB}$ in the isolated resonators, which are split either in frequency ($AB>0$) or in life time ($AB<0$), but not in both.
For a numerical example of coupled resonators with real $A$, $B$ and $W$ see Ref.~\cite{PhysRevA.90.053819}, where this situation was realized in a coupled-resonator chain of circular disks with attached nanoparticles.

In summary, the general model presented in this work requires reciprocity and evanescent coupling at smooth interfaces. The variant with an additional $\mathcal{PT}$-symmetry requires to tune a single parameter, which aligns the resonances of an isolated resonator either in frequency or life time.

\section{Wave matching}
\label{appB}
Here, we present the details of the derivation of the edge-state conditions
\eqref{symmmain}-\eqref{lambdamain}. These are obtained by wave matching from the quasi-one-dimensional tight-binding equations
\eqref{eq:cmeqs1d} with the interface defined by Eq.~\eqref{eq:hn},
which we first carry out in $(A',B')$ parameter space.
The edge-state arcs in $k$ space and in $(\mathcal{A},\mathcal{B})$ space then follow from the parameterizations \eqref{abprime} and \eqref{effective}.

To carry out this wavematching, we first determine the Bloch solutions
\begin{equation}
\boldsymbol{\varphi}_{n}=\lambda^n\boldsymbol{\Phi}
\end{equation}
in the regions on both sides of the interface for
a given value of $\Omega$, where we utilize propagation factors $\lambda=\exp(ik_x)$. According to Eq.~\eqref{eq:cmeqs1d}, the Bloch states in the region $n<0$ are given by the condition
\begin{equation}
\Omega\boldsymbol{\Phi}
=
        \begin{pmatrix}
0 & A^{\prime} \\
B^{\prime}  & 0
\end{pmatrix}
\boldsymbol{\Phi}
+
(\lambda+\lambda^{-1})W_x
        \begin{pmatrix}
0 & 1 \\
1  & 0
\end{pmatrix}
\boldsymbol{\Phi}.
\end{equation}
This permits nontrivial solutions if
\begin{equation}
\left|\left|\begin{array}{cc}
-\Omega & A'+2W_xC \\
 B'+2W_xC & -\Omega
\end{array}\right|\right|=0,
\end{equation}
where we introduced $\lambda+\lambda^{-1}=2 C$.
The two solutions of the resulting quadratic equation are
\begin{subequations}\label{eq:c}%
\begin{eqnarray}
&&2W_xC_1=
-\frac{A'+B'}{2} + \sqrt{\frac{(A'-B')^{2}}{4}+\Omega^{2}},
\\
&&2W_xC_2=
-\frac{A'+B'}{2} - \sqrt{\frac{(A'-B')^{2}}{4}+\Omega^{2}},
\end{eqnarray}
\end{subequations}
which corresponds to four propagation factors
\begin{subequations}\label{lambda}%
\begin{eqnarray}
\lambda_{1}^{(\pm)}= C_{1} \pm \sqrt{C_1^2-1}\equiv C_1\pm S_1,\\
\lambda_{2}^{(\pm)}= C_{2} \pm \sqrt{C_2^2-1}\equiv C_2\pm S_2.
\end{eqnarray}%
\end{subequations}
We note that the associated Bloch vectors
\begin{equation}
\boldsymbol{\Phi}_l^{(\rm L)}\propto\begin{pmatrix}A^{\prime}+2W_xC_{l}\\ \Omega\end{pmatrix}
\label{eq:blochl}
\end{equation}
are the same for each pair $\lambda_l^{(+)}$ and $\lambda_l^{(-)}$, and that $\lambda_l^{(+)}\lambda_l^{(-)}=1$, which are both consequences of reciprocity.
The same construction can be carried out on the right side of the interface, where one obtains the same propagation factors but associated with the
Bloch vectors
\begin{equation}
\boldsymbol{\Phi}_l^{(\rm R)}=\sigma_x\boldsymbol{\Phi}_l^{(\rm L)}\propto\begin{pmatrix}\Omega\\ A^{\prime}+2W_xC_{l}\end{pmatrix}.
\label{eq:blochr}
\end{equation}

Superpositions of these Bloch waves form the general solutions to the left and right of the interface. Edge states are obtained when we can match solutions that decay to both sides of the interface. This requires to depart from the bands, hence to study values of $\Omega$ where the propagation factors are no longer unimodular. To be specific, we select from each pair $\lambda_l^{(+)},\lambda_l^{(-)}$
the propagation factor $\lambda_l$ fulfilling
\begin{equation}
|\lambda_l|> 1,
\label{eq:lcondition}
\end{equation}
and from here on work with the propagation factors $\lambda_l, 1/\lambda_l$.
(By a proper choice of the branch of the square roots in Eq.~\eqref{lambda}, we could alternative enforce that always $|\lambda_1^{(+)}|, |\lambda_2^{(+)}|>1$.) The assumption \eqref{eq:lcondition}
will be revisited below to confirm the consistency of a potential edge state.

We first deal with the case of solutions that are symmetric about the interface (the antisymmetric case will follow from chiral symmetry).
These are of the form
\begin{eqnarray}
&&\psi_{n<0}= a_1\lambda^{n+1}_{1} \boldsymbol{\Phi}_1^{(\rm L)}
+ a_2\lambda_{2}^{n+1} \boldsymbol{\Phi}_2^{(\rm L)},
\\
&&\psi_{n\geq 0}= a_1\lambda^{-n}_{1}\boldsymbol{\Phi}_1^{(\rm R)}
+ a_2\lambda^{-n}_{2} \boldsymbol{\Phi}_2^{(\rm R)}.
\label{eq:edgesym}
\end{eqnarray}

The consistency of these expressions across the interface  can be read off by comparing the expressions at $n=0$, or equivalently those at $n=-1$, which in both cases delivers the wave-matching condition
\begin{eqnarray}
a_1\lambda_{1}\boldsymbol{\Phi}_1^{(\rm L)}
+ a_2\lambda_{2}\boldsymbol{\Phi}_2^{(\rm L)}
= a_1 \boldsymbol{\Phi}_1^{(\rm R)}
+ a_2\boldsymbol{\Phi}_2^{(\rm R)}.
\label{eq:edgesymmatch}
\end{eqnarray}

This has nontrivial solutions if
\begin{align}
&\left|\left|
\begin{array}{cc}
A'+2W_xC_1-\lambda_{1}\Omega
&
A'+2W_xC_2-\lambda_{2}\Omega\\
\Omega-\lambda_{1}(A'+2W_xC_1)
&
\Omega-\lambda_{1}(A'+2W_xC_1)
\end{array}
\right|\right|=0
\\
\Rightarrow&
W_x(1-\lambda_{1}\lambda_{2})(C_1-C_2)=\Omega(\lambda_{1}-\lambda_{2})
\\
\Rightarrow&
2-\lambda_{1}\lambda_{2}-\frac{1}{\lambda_{1}\lambda_{2}}=\frac{2\Omega}{W_x},
\label{eq:consistency}
\end{align}
where in the first step we used $(A'+2W_xC_1)(A'+2W_xC_2)=-\Omega^2$ and in the second step we multiplied through by $1-1/(\lambda_{1}\lambda_{2})$.

Using the definitions introduced in Eq.~\eqref{lambda}, condition \eqref{eq:consistency}
 can be rewritten as
\begin{equation}
1-C_1C_2-\Omega/W_x=\pm S_1S_2,
\label{eq:ccss}
\end{equation}
where the sign depends on which of the propagation factors fulfill the condition \eqref{eq:lcondition}. Squaring both sides and using the identities $W_x^2C_1C_2=(A'B'-\Omega^2)/4$,
  $W_x^2(C_1-C_2)^2=\Omega^2+(A'-B')^2/4$, we find the condition
\begin{equation}
\Omega(\Omega-2W_{x})^{2} -A^{\prime}B^{\prime}\Omega + (A^{\prime}-B^{\prime})^{2}\frac{W_{x}}{2} =0,
\label{symm}
\end{equation}
which must be fulfilled for any edge state with a symmetric mode profile.

The corresponding condition for edge states with an antisymmetric mode profile can be constructed by envoking the chiral symmetry operation $\sigma_z$, which transforms the Bloch vectors
given in Eqs.~\eqref{eq:blochl} and \eqref{eq:blochr} according to
\begin{equation}
\sigma_z\boldsymbol{\Phi}_l^{(\rm R)}=\sigma_z\sigma_x\boldsymbol{\Phi}_l^{(\rm L)}=
-\sigma_x\sigma_z\boldsymbol{\Phi}_l^{(\rm L)}.
\end{equation}
As the chiral symmetry inverts the sign of the complex frequency $\Omega$, this gives
\begin{align}
&2-\lambda_{1}\lambda_{2}-\frac{1}{\lambda_{1}\lambda_{2}}=-\frac{2\Omega}{W_x}
\label{eq:consistency2}
\\
\Rightarrow &
-\Omega(\Omega+2W_{x})^{2} +A^{\prime}B^{\prime}\Omega + (A^{\prime}-B^{\prime})^{2}\frac{W_{x}}{2} =0,
\label{antisymm}
\end{align}
which must be fulfilled for any edge state with an antisymmetric mode profile.

Equations \eqref{symmmain} and \eqref{antisymmmain} follow by first expressing Eqs.~\eqref{symm} and \eqref{antisymm} according to the definitions \eqref{abprime} as
\begin{align}
&(A+2W_{y}\cos{k_{y}})(B+2W_{y}\cos{k_{y}})\Omega
\nonumber\\
&=\Omega(\Omega-2W_{x})^{2}  + (A-B)^{2}\frac{W_{x}}{2}
\label{symmmain2}
\end{align}
and
\begin{align}
&(A+2W_{y}\cos{k_{y}})(B+2W_{y}\cos{k_{y}})\Omega
\nonumber\\
&=\Omega(\Omega+2W_{x})^{2} - (A-B)^{2}\frac{W_{x}}{2}
,
\label{antisymmmain2}
\end{align}
and then converting these to the effective parameters introduced in Eq.\ \eqref{effective}, which amount to $\mathcal{A}=(A'+B')/2W_x$, $\mathcal{B}=(A'-B')/2W_x$.
The propagation factors \eqref{lambdamain} follow analogously from Eqs.~\eqref{eq:c} and \eqref{lambda} with the proper branch of the square root, and the consistency conditions \eqref{eq:consistencymain} and \eqref{eq:consistency2main} follow from Eqs.~\eqref{eq:consistency} and \eqref{eq:consistency2}.

The physical interpretation of these consistency conditions is as follows.
In order for a solution $\Omega$ of Eqs.~\eqref{symm} or Eqs.~\eqref{antisymm} to define an actual edge state, the corresponding  wave function must decay away from the interface. This requires that $\Omega$ inserted into Eq.~\eqref{eq:c} gives propagation factors
$|\lambda_{1}|$, $|\lambda_{2}|>1$ according to our choice in Eq.~\eqref{eq:lcondition},
but the information about this choice is lost when one squares Eq.~\eqref{eq:ccss}.
If for one of the two propagation factors we find $|\lambda_{l}|=1$, we instead obtain a scattering state. If the consistency equations \eqref{eq:consistency} and \eqref{eq:consistency2} can only be fulfilled by combining a decaying and an increasing branch of $\lambda^{(\pm)}_{l}$, the state is unphysical. Therefore, in ($A',B'$) parameter space, solutions change from physical to unphysical when for one pair of propagation factors $|\lambda^{(\pm)}_{l}|=1$. This corresponds to propagating waves in the bulk system, and therefore occurs when an edge-state arc meets the bulk dispersion relation.

In the $\mathcal{PT}$-symmetric case, the conditions where the arcs meet the bands can be determined analytically from the realization that they coincide with exceptional points where the frequencies $\Omega$ of two edge states become degenerate, with an either real or imaginary value. Demanding degeneracy of solutions in the cubic equations
Eq.~\eqref{symm} and \eqref{antisymm},
we find the condition
\begin{equation}
\frac{27}{8}(A'+B')^{4}=2A^{\prime 2}B^{\prime 2}(1+\frac{A'B'}{W_x^{2}})+(8W_x^{2}+9A'B')(A'+B')^{2}
\end{equation}
for the termination points of the arcs on the real branches of the band structure,
and the condition
\begin{equation}
A^{\prime 2}+6A'B'+B^{\prime 2}=32W_x^2
\end{equation}
for their termination points on the imaginary branches of the band structure.
In Fig.~\ref{fig4}(a), these conditions are again translated from $(A',B')$ space into $(\mathcal{A},\mathcal{B})$ space.

\end{document}